\documentclass[conference,9pt]{IEEEtran}
\IEEEoverridecommandlockouts
\usepackage{array}
\usepackage{cite}
\usepackage{amsmath,amssymb,amsfonts}
\usepackage{algorithmic}
\usepackage{graphicx}
\usepackage{textcomp}
\usepackage{xcolor}
\usepackage{hyperref}

\usepackage{tikz}
\usepackage{pgfplots}
\def\BibTeX{{\rm B\kern-.05em{\sc i\kern-.025em b}\kern-.08em
    T\kern-.1667em\lower.7ex\hbox{E}\kern-.125emX}}
\begin{document}

\title{Improving Speech Emotion Recognition in Under-Resourced Languages via Speech-to-Speech Translation with Bootstrapping Data Selection}

\renewcommand{\footnoterule}{
  \kern -3pt
  \hrule width \linewidth height 0.5pt
  \kern 2.5pt
}
\DeclareRobustCommand*{\IEEEauthorrefmark}[1]{
  \raisebox{0pt}[0pt][0pt]{\textsuperscript{\footnotesize #1}}
}
\author{
    \IEEEauthorblockN{
        Hsi-Che Lin\IEEEauthorrefmark{1}\textsuperscript{\textasteriskcentered},
        Yi-Cheng Lin\IEEEauthorrefmark{1}\textsuperscript{\textasteriskcentered},
        Huang-Cheng Chou\IEEEauthorrefmark{2} and
        Hung-yi Lee\IEEEauthorrefmark{1}
    }
    \vspace{3pt}
    \IEEEauthorblockA{
        \IEEEauthorrefmark{1}National Taiwan University, Taipei, Taiwan\\
        \IEEEauthorrefmark{2}Department of Electrical Engineering at National Tsing Hua University, Hsinchu, Taiwan\\
        Email: 
            \{r13942079, r12942075\}@ntu.edu.tw, 
            huangchengchou@gmail.com,
            hungyilee@ntu.edu.tw
    }
    \vspace{-20pt}
    \thanks{\textsuperscript{\textasteriskcentered} Equal contribution.}
}

\maketitle
\begin{abstract}
Speech Emotion Recognition (SER) is a crucial component in developing general-purpose AI agents capable of natural human-computer interaction. However, building robust multilingual SER systems remains challenging due to the scarcity of labeled data in languages other than English and Chinese. In this paper, we propose an approach to enhance SER performance in low SER resource languages by leveraging data from high-resource languages. Specifically, we employ expressive Speech-to-Speech translation (S2ST) combined with a novel bootstrapping data selection pipeline to generate labeled data in the target language. Extensive experiments demonstrate that our method is both effective and generalizable across different upstream models and languages. Our results suggest that this approach can facilitate the development of more scalable and robust multilingual SER systems. Our code is available at: \url{https://github.com/hsi-che-lin/Improve-SER-via-S2ST}
\end{abstract}

\begin{IEEEkeywords}
multilingual, emotion, speech translation
\end{IEEEkeywords}

\section{Introduction}

Speech Emotion Recognition (SER) is key to improving human-computer interaction (HCI) \cite{ramakrishnan2013speech, alsabhan2023human} by enabling machines to understand and respond to users' emotions, enhancing experiences in areas like customer service, mental health, and virtual assistants. While large SER datasets exist for high-resource languages like English \cite{msp_podcast}, Chinese \cite{biic_podcast}, and Russian \cite{dusha}, many languages have far smaller datasets, often under 10 hours of speech, hindering SER performance. Addressing this gap is critical for developing inclusive emotion recognition systems across diverse linguistic communities. 
As a result, our research question is \textbf{whether it is possible to boost the performance of SER on under-resourced languages}, which, in this paper, refer to languages lacking large labeled SER datasets.

Several approaches have been investigated to address this problem, which can be broadly categorized into three groups: transfer learning, domain adaptation, and data augmentation.
Transfer learning methods \cite{neumann2018cross, AHMAD2020112851} aim to enhance performance in low-resource languages by leveraging models trained on high-resource languages.
However, when the amount of data available in different languages is imbalanced, a model predominantly trained on a high-resource language might become biased towards the features of that language, neglecting the characteristics of the lower-resource languages.
Domain adaptation methods focus on aligning features across languages through techniques such as feature normalization \cite{hozjan2003context, sagha16_interspeech}, direct alignment \cite{song2016cross, hassan2013acoustic, 10095388}, and domain adversarial learning \cite{10095388}. Despite these efforts, feature space alignment does not always ensure accurate prediction alignment, as certain undesired information may still be encoded in the features.
For augmentation based methods, people usually adopt GAN \cite{chatziagapi19_interspeech, eskimez20_interspeech, latif20_interspeech,lu19_adan,latif2023generative} and CycleGAN \cite{bao19_interspeech, su19_conditional_cyclegan} types of methods.
However, instability training of GAN makes generating high-quality data difficult and limits their usability.

A recent study \cite{ma2024leveraging} leverages GPT-4 \cite{gpt4technicalreport} and expressive Text-to-Speech (TTS) for data augmentation, enabling the generation of large datasets. However, the synthesized data is conditioned on predefined text and discrete emotion classes, which lack naturalness, fail to capture nuanced paralinguistic cues such as hesitations \cite{li2024spontts} and vocal burst \cite{Brooks_2023}, and overlook the possibility of multiple, complex emotions co-occurring within a single speech instance \cite{Cowen_2021,Chou_2024}. Moreover, it depends on Azure TTS, which supports only English, restricting its applicability to other languages.

In this paper, we generate synthetic target language data directly from data of a high-resource language using expressive S2ST. This approach offers three benefits: it avoids feature-prediction mismatches in domain adaptation method by working directly in the speech sample space, addresses language imbalance in transfer learning through target data generation, and captures more paralinguistic cues and emotional nuances by conditioning on real-world samples.

\begin{figure}
\centering
\includegraphics[width=1.0\columnwidth]{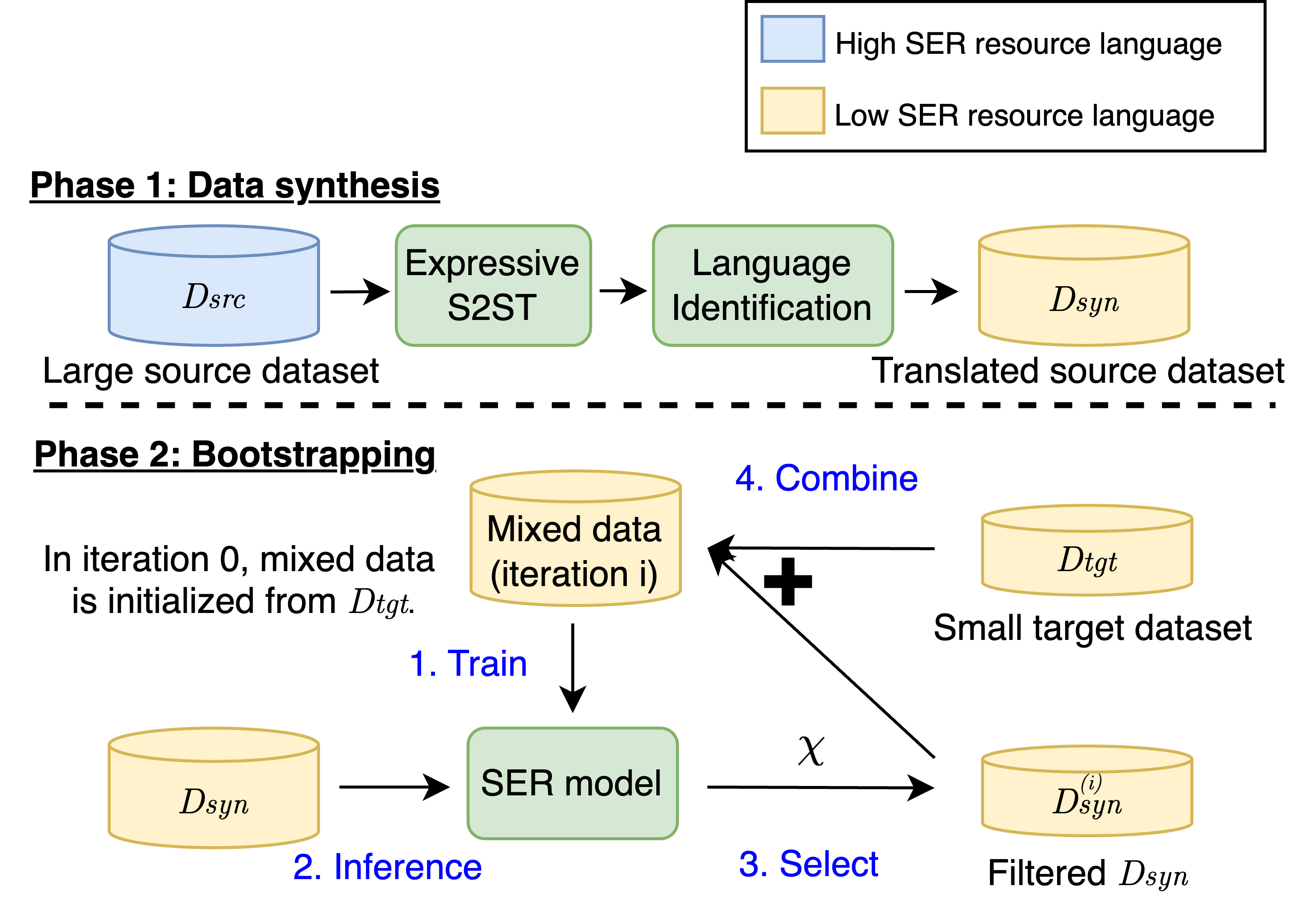}
\vspace{-20pt}
\caption{Overview of our method.}
\vspace{-15pt}
\label{fig:overview}
\end{figure}

As shown in Fig.\ref{fig:overview}, our method uses a two-phase pipeline: data synthesis and bootstrapping data selection. First, we generate target language data using an expressive S2ST model. Then, we apply a novel bootstrapping method to iteratively select the most beneficial data for training according to the prediction of the model from previous iterations. Despite its simplicity, our method consistently improves performance across various models, languages, and datasets.

Our work yields the following contributions:
\vspace{-6pt}
\begin{itemize}
    \item We show that it is possible to improve SER in under-resourced languages by expressive S2ST and large-scale datasets of high-resource language.
    \item We propose a simple, effective method to achieve consistent improvement in different upstream models, languages, and datasets.
    \item Based on our experiment, we give some principles for choosing high-resource datasets that can achieve consistent improvement.
\end{itemize}

{\begin{table}[]
\setlength\tabcolsep{4.75pt}
\renewcommand{\arraystretch}{1.15}
\caption{Overview of datasets involved in our experiments. The first two rows are used as source datasets, and the rest are used as target datasets. We list the number of speakers, utterances, and duration to highlight their difference in scale.}
\vspace{-10pt}
\begin{center}
\begin{tabular}{c|ccccc}
\hline
\textbf{Dataset} & \textbf{Type} & \textbf{Language} & \textbf{\#Spk}$^{\mathrm{a}}$ & \textbf{\#Utts}$^{\mathrm{b}}$ & \textbf{\#Hrs}$^{\mathrm{c}}$ \\ \hline
MSP-Podcast \cite{msp_podcast}     & Podcast         & English       & 2172+         & 149307           & 235.94          \\
BIIC-Podcast \cite{biic_podcast}    & Podcast         & Chinese       & Unknow             & 70000                & 147.43               \\ \hline
EmoDB \cite{emodb}           & Act             & German        & 10           & 339             & 0.26           \\
CaFE \cite{cafe}            & Act             & French        & 12           & 843             & 0.62           \\
EMOVO \cite{emovo}           & Act             & Italian       & 6            & 1179            & 0.28           \\
MESD \cite{mesd}            & Act             & Spanish       & 11           & 1753            & 0.11           \\ \hline
\multicolumn{6}{l}{Number of $^{\mathrm{a}}$speakers, $^{\mathrm{b}}$utterances, $^{\mathrm{c}}$hours}
\end{tabular}
\vspace{-20pt}
\end{center}
\label{tab:data}
\end{table}
}

\section{Method}
\vspace{-6pt}
\subsection{Problem definition}
\vspace{-2pt}
This paper aims to enhance the performance of SER on a small target dataset, $D_{tgt}$, in a target language $L_t$.
We assume access to a large-scale SER dataset, $D_{src}$, in a source language $L_s$, which differs from $L_t$.
All samples in both datasets are assumed to have four-class emotion labels, that is, \textit{angry}, \textit{happy}, \textit{neutral}, and \textit{sad}.
Note that, in this paper, we do not consider unlabeled data that may be used in other settings like semi-supervised learning \cite{agarla2024semi, parthasarathy2020semi, zhang2016enhanced}.

\vspace{-4pt}
\subsection{Phase 1 - Data synthesis}
\vspace{-2pt}
Given a SER dataset in a high-resource source language ($D_{src}$), we propose translating each sample into the target language $L_t$ using an expressive S2ST model, thereby creating a translated version of the source dataset.
We observe the S2ST system sometimes may fail at translating. One common failure mode is that the output language is not the target language. As a result, we apply a language identification model, Whisper Large v3 \footnote[1]{https://github.com/openai/whisper} \cite{whisper}, to filter out samples detected as non-target language.
The resulting filtered dataset, \( D_{syn} = \{(x_i, y_i)\}_{i=0}^N \), where $x_i$ and $y_i$ represent a speech sample and its label, respectively, and $N$ is the number of samples in $D_{src}$, is used for phase $2$.

\vspace{-4pt}
\subsection{Phase 2 - Bootstrapping}
\vspace{-2pt}
After some pilot studies, we find that naively using all $D_{syn}$ for training often leads to great performance degradation.
To address this, we propose a bootstrapping method that iteratively selects data from $D_{syn}$ more likely to be from a distribution similar to $D_{tgt}$.

Let $M_i$ represent the model trained on $D^{(i)}_{syn} \cup D_{tgt}$ in the $i$-th iteration. We define a criterion, $\chi$, which takes a data sample and its corresponding model prediction as input, and outputs either $0$ or $1$. The selected synthesized data for iteration $i+1$ is given by
\begin{equation}
    D^{(i+1)}_{syn}=\{(x, y) \in D_{syn} \;|\; \chi(x, y, M_i(x)) = 1\}\text{.}
\end{equation}
We let $D^{(0)}_{syn}=\phi$, that is, $M_0$ is trained solely on $D_{tgt}$.
The total number of iterations $I$ is a hyperparameter, defaulting to $2$, and we use $M_I$ for final evaluation.

\vspace{-4pt}
\subsection{Selection criterion for bootstrapping}
\vspace{-2pt}
In this paper, we propose two selection criteria, denoted as $\chi$. We hypothesize that the performance drop observed when naively using all $D_{syn}$ is due to a domain discrepancy between $D_{tgt}$ and $D_{syn}$.
Our intuition is that if the baseline model can predict a synthesized data sample with reasonable accuracy, that sample likely originates from a distribution similar to that of the target data\cite{zheng2020out}.
Therefore, our first selection criterion retains samples that are correctly predicted by the model from the previous iteration. That is
\begin{align}
\chi(x,y,M) =
\begin{cases} 
1 & \text{if } \arg\max M(x) = y \\
0 & \text{otherwise.}
\end{cases}
\label{eq:1}
\end{align}

Additionally, given the availability of soft labels—a probability distribution derived from the voting results of multiple annotators and adjusted using a smoothing technique\cite{szegedy2016rethinking}—we leverage these labels to account for the multidimensional nature of human emotions\cite{Cowen_2021}.
This forms the basis of our second selection criterion, which we adopt as the default approach.
The criterion is given by
\begin{align}
\chi(x,y,M) =
\begin{cases} 
1 & \begin{aligned} 
       & \text{if } \arg\max M(x) = \arg\max y \\
       & \text{and } KL\left(M(x),y\right) < \operatorname{med}_{M,D_{syn}}
    \end{aligned} \\[7pt]
0 & \text{otherwise,}
\end{cases}
\label{eq:2}
\end{align}
where $y$ here is a distribution rather a label ID as it is in equation \eqref{eq:1}, $KL$ stands for Kullback–Leibler (KL) divergence between two distribution, and $\operatorname{med}_{M,D_{syn}}$ means the median of KL divergence of all prediction and soft labels in $D_{syn}$.

\vspace{-3pt}
\section{Experiment setup}
\vspace{-4pt}
\subsection{Datasets}
\vspace{-2pt}
The overview of the datasets utilized in this paper is shown in Tab.~\ref{tab:data}.
For high-resource SER datasets $D_{src}$, we choose to use MSP-Podcast \cite{msp_podcast}, a large-scale English dataset, by default, and we also experiment with a large-scale Chinese dataset, BIIC-Podcast \cite{biic_podcast}, for comparison.
For target datasets, we experiment with four target datasets $D_{tgt}$ of different languages $L_t$, including EmoDB \cite{emodb}, CaFE \cite{cafe}, EMOVO \cite{emovo}, and MESD \cite{mesd}, which correspond to German, French, Italian, and Spanish, respectively.

\vspace{-2pt}
\subsection{Speech Translation}
\vspace{-2pt}
We use Seamless Expressive\footnote[2]{https://github.com/facebookresearch/seamless\_communication} \cite{seamless} as the expressive speech translation model.
We also use S2ST models trained without expressivity, SeamlessM4T-Large v2, for comparison.
Seamless Expressive is built upon SeamlessM4T-Large v2 as its foundational model and integrates an expressivity encoder to guide the generation of speech units, ensuring appropriate rhythm, speaking rate, and pauses.
Additionally, it replaces the HiFi-GAN \cite{hifigan} vocoder with an expressive unit-to-speech generator to convey vocal styles more effectively. We use a speech unit generation beam size of 5.

{\begin{table}[]
\setlength\tabcolsep{5.25pt}
\renewcommand{\arraystretch}{1.15}
\caption{Overview of the upstream model in our experiments. We highlight the language and whether emotional data or paralinguistic-aware objectives are used during pretraining.}
\vspace{-10pt}
\begin{center}
\begin{tabular}{c|cccc}
\hline
\textbf{}    & \textbf{Language} & \textbf{Emotion} & \textbf{\#Param.} & \textbf{Hidden dim.} \\ \hline
emotion2vec \cite{emotion2vec}  & Eng.              & $\checkmark$     & 93.8M              & 768                  \\
exp. encoder$^{\mathrm{a}}$ \cite{seamless} & Multi.            & $\checkmark$     & 7.2M               & 512                  \\
WavLM \cite{wavlm}       & Eng.              &                  & 316.6M             & 1024                 \\
XLS-R 300M \cite{xls_r}  & Multi.            &                  & 315.4M             & 1024                 \\ \hline
\multicolumn{5}{l}{$^{\mathrm{a}}$ Expressivity encoder in Seamless Expressive}
\end{tabular}
\vspace{-20pt}
\label{tab:model}
\end{center}
\end{table}
}
{\begin{table*}[]
\setlength\tabcolsep{5.25pt}
\renewcommand{\arraystretch}{1.15}
\caption{Result of applying our method to different upstream models and target datasets.}
\vspace{-10pt}
\begin{center}
\begin{tabular}{p{50pt}|ccc|ccc|ccc|ccc}
\hline
\multicolumn{1}{c|}{}      & \multicolumn{3}{c|}{\textbf{EmoDB (de)}}                                                            & \multicolumn{3}{c|}{\textbf{CaFE (fr)}}                                                                      & \multicolumn{3}{c|}{\textbf{EMOVO (it)}}                                                                     & \multicolumn{3}{c}{\textbf{MESD (es)}}                                                                       \\ \cline{2-13} 
\multicolumn{1}{c|}{}      & \textbf{UA(\%)$\uparrow$} & \textit{\textbf{WA(\%)$\uparrow$}} & \textit{\textbf{F1(\%)$\uparrow$}} & \textit{\textbf{UA(\%)$\uparrow$}} & \textit{\textbf{WA(\%)$\uparrow$}} & \textit{\textbf{F1(\%)$\uparrow$}} & \textit{\textbf{UA(\%)$\uparrow$}} & \textit{\textbf{WA(\%)$\uparrow$}} & \textit{\textbf{F1(\%)$\uparrow$}} & \textit{\textbf{UA(\%)$\uparrow$}} & \textit{\textbf{WA(\%)$\uparrow$}} & \textit{\textbf{F1(\%)$\uparrow$}} \\ \hline
emotion2vec                & 84.97                     & 83.79                              & 84.04                              & 67.33                              & 66.72                              & 66.02                              & 49.90                              & 49.90                              & 44.22                              & 64.75                              & 65.18                              & 64.33                              \\
\multicolumn{1}{r|}{+ours} & \textbf{85.94}            & \textbf{85.17}                     & \textbf{85.27}                     & \textbf{68.32}                     & \textbf{68.29}                     & \textbf{67.74}                     & \textbf{55.56}                     & \textbf{55.56}                     & \textbf{51.04}                     & \textbf{64.87}                     & \textbf{65.21}                     & \textbf{64.65}                     \\ \hline
exp. encoder$^{\mathrm{a}}$               & 92.73                     & 92.41                              & 92.56                              & 64.55                              & 62.62                              & 63.11                              & 53.37                              & 53.37                              & 49.20                              & 84.09                              & 84.08                              & 84.03                              \\
\multicolumn{1}{r|}{+ours} & \textbf{93.43}            & \textbf{93.01}                     & \textbf{93.30}                     & \textbf{65.01}                     & \textbf{63.31}                     & \textbf{64.34}                     & \textbf{54.27}                     & \textbf{54.27}                     & \textbf{49.44}                     & \textbf{84.49}                     & \textbf{84.23}                     & \textbf{84.42}                     \\ \hline
WavLM large                & 93.65                     & 94.54                              & 93.74                              & 74.80                              & 74.19                              & 74.45                              & 58.33                              & 58.33                              & 53.36                              & 68.00                              & 68.69                              & 67.34                              \\
\multicolumn{1}{r|}{+ours} & \textbf{94.11}            & \textbf{94.94}                     & \textbf{94.17}                     & \textbf{76.12}                     & \textbf{75.64}                     & \textbf{76.18}                     & \textbf{64.98}                     & \textbf{64.98}                     & \textbf{62.21}                     & \textbf{68.12}                     & \textbf{68.71}                     & \textbf{67.96}                     \\ \hline
XLS-R 300M                 & 80.68                     & 79.17                              & 79.67                              & 44.18                              & 40.80                              & 37.97                              & 46.43                              & 46.43                              & 40.25                              & 73.93                              & 74.18                              & 73.70                              \\
\multicolumn{1}{r|}{+ours} & \textbf{81.83}            & \textbf{80.60}                     & \textbf{80.74}                     & \textbf{51.72}                     & \textbf{48.61}                     & \textbf{49.67}                     & \textbf{47.42}                     & \textbf{47.42}                     & \textbf{41.91}                     & \textbf{77.00}                     & \textbf{77.71}                     & \textbf{76.73}                     \\ \hline
\multicolumn{13}{l}{$^{\mathrm{a}}$Expressivity encoder in Seamless Expressive}
\end{tabular}
\vspace{-20pt}
\end{center}
\label{tab:main}
\end{table*}
}

{\begin{table*}[]
\setlength\tabcolsep{5.25pt}
\renewcommand{\arraystretch}{1.15}
\caption{Comparison to other data augmentation methods.}
\vspace{-10pt}
\begin{center}
\begin{tabular}{p{50pt}|ccc|ccc|ccc|ccc}
\hline
\multicolumn{1}{c|}{\textbf{}}          & \multicolumn{3}{c|}{\textbf{EmoDB (de)}}                                                                     & \multicolumn{3}{c|}{\textbf{CaFE (fr)}}                                                                      & \multicolumn{3}{c|}{\textbf{EMOVO (it)}}                                                                     & \multicolumn{3}{c}{\textbf{MESD (es)}}                                                                       \\ \cline{2-13} 
\multicolumn{1}{c|}{\textit{\textbf{}}} & \textit{\textbf{UA(\%)$\uparrow$}} & \textit{\textbf{WA(\%)$\uparrow$}} & \textit{\textbf{F1(\%)$\uparrow$}} & \textit{\textbf{UA(\%)$\uparrow$}} & \textit{\textbf{WA(\%)$\uparrow$}} & \textit{\textbf{F1(\%)$\uparrow$}} & \textit{\textbf{UA(\%)$\uparrow$}} & \textit{\textbf{WA(\%)$\uparrow$}} & \textit{\textbf{F1(\%)$\uparrow$}} & \textit{\textbf{UA(\%)$\uparrow$}} & \textit{\textbf{WA(\%)$\uparrow$}} & \textit{\textbf{F1(\%)$\uparrow$}} \\ \hline
baseline                                & 84.97                              & 83.79                              & 84.04                              & 67.33                              & 66.72                              & 66.02                              & 49.90                              & 49.90                              & 44.22                              & 64.75                              & 65.18                              & 64.33                              \\
noise                                   & 84.57                              & 84.07                              & 84.10                              & 63.29                              & 63.14                              & 61.46                              & 49.80                              & 49.80                              & 45.01                              & 64.69                              & 65.13                              & 64.55                              \\
CopyPaste                               & 84.66                              & 83.93                              & 84.34                              & 66.80                              & 65.91                              & 65.43                              & 55.12                              & 55.12                              & 50.48                              & \textbf{64.87}                     & \textbf{65.37}                     & \textbf{64.69}                     \\
Ours                                    & \textbf{85.94}                     & \textbf{85.17}                     & \textbf{85.27}                     & \textbf{68.32}                     & \textbf{68.29}                     & \textbf{67.74}                     & \textbf{55.56}                     & \textbf{55.56}                     & \textbf{51.04}                     & \textbf{64.87}                     & 65.21                              & 64.65 \\ \hline
\end{tabular}
\vspace{-20pt}
\label{tab:compare}
\end{center}
\end{table*}
}

\vspace{-2pt}
\subsection{Upstream model}
\vspace{-2pt}
To demonstrate the generalizability of our method, we experiment with four different upstream models: emotion2vec \cite{emotion2vec}, WavLM \cite{wavlm}, Wav2Vec2 XLS-R 300M \cite{xls_r}, and the expressivity encoder from Seamless Expressive \cite{seamless}.
The overview of these models is shown in Tab.~\ref{tab:model}.
These models were selected for their strong SER performance \cite{emobox, emotion2vec,Wu_2024} and diverse pretraining conditions. Some models are trained on multilingual data, others on English, and some use emotional data or paralinguistic-aware objectives during pretraining, while others do not.
We selected the expressivity encoder from Seamless Expressive, which is not a common choice for SER upstream models, not only because it is a multilingual, paralinguistic-aware pre-trained model, but also because it plays a key role in our data synthesis pipeline. This connection may help explain the effectiveness of our method.

\subsection{Evaluation protocol}
\vspace{-3pt}
Our evaluation protocol primarily follows the EmoBox framework \cite{emobox}.
We perform speaker-aware cross-validation, ensuring that speakers in the training and test sets are different.
The partitioning of folds in each dataset also follows EmoBox.
To reduce performance variance, we repeat each experiment with three different random seeds and average the results, as we observed sensitivity to initialization, particularly on smaller target datasets.

The downstream model is a simple two-layer feed-forward network with ReLU \cite{relu} activation, incorporating a pooling layer between the layers.
The inputs to this network are the final layer features extracted by the upstream model.
In all experiments, the upstream model is frozen, and the downstream model is trained using the standard cross-entropy loss function.
For each experiment, we report three commonly used performance metrics: Unweighted Average Accuracy (UA), Weighted Average Accuracy (WA), and Macro F1 Score (F1).

\vspace{-2pt}
\section{Result and Discussion}
\vspace{-2pt}
\subsection{Main result}
\vspace{-2pt}
As shown in Table~\ref{tab:main}, our proposed method achieves performance improvements across all upstream models and target datasets, regardless of language.
Notably, performance gains are observed even on well-performing datasets like EMO-DB, while on underperforming datasets, such as WavLM on EMOVO or XLS-R on CaFE, our method boosts the F1 score by as much as 8 to 11 points.

Our experiments include models pretrained in English or multilingual settings and models pretrained for prosody or emotion awareness.
The consistent improvements demonstrate the generalizability of our method across various types of upstream models. Importantly, even on a multilingual and prosody-aware pretrained model (the expressivity encoder in Seamless Expressive), we observe performance gains, underscoring the effectiveness of scaling up the downstream training dataset.

Interestingly, we find that the expressivity encoder in Seamless Expressive, which, to our knowledge, has not been previously used as an upstream model for SER tasks, performs remarkably well despite its smaller model size and reduced hidden dimensions.
We attribute this to its strong ability to capture subtle paralinguistic cues that are closely correlated with emotional information—an essential factor in Seamless Expressive’s ability to generate high-quality expressive speech.
Since our method relies on high-quality data synthesized by Seamless Expressive, the strong emotion recognition performance of Seamless Expressive further explains the effectiveness of our approach.
We validate this further through an ablation study, where we swap Seamless Expressive with its non-expressive counterpart, as detailed in the next section.

\vspace{-3pt}
\subsection{Comparison to other methods}
\vspace{-3pt}
Since our method can be considered a type of data augmentation, we compare it with two other data augmentation methods—noise augmentation and CopyPaste \cite{pappagari2021copypaste}—to demonstrate its effectiveness.
As shown in Table~\ref{tab:compare}, while CopyPaste performs slightly better on one dataset, only our method consistently improves performance across all languages.
Notably, the upstream models used already exhibit strong performance in multilingual ER \cite{emobox}, making consistent improvement challenging even for SOTA augmentation methods like CopyPaste.
This further highlights the effectiveness of our proposed method.

{\begin{table*}[h]
\newcolumntype{P}[1]{>{\centering\arraybackslash}p{#1}}
\setlength\tabcolsep{3.75pt}
\renewcommand{\arraystretch}{1.15}
\caption{Ablation study with data synthesis and selection process}
\vspace{-10pt}
\begin{center}
\begin{tabular}{P{12pt}|P{18pt}@{}P{18pt}@{}P{18pt}@{}P{20pt}@{}|ccc|ccc|ccc|ccc}
\hline
                      & \textbf{}            & \textbf{}            & \textbf{}            & \textbf{}            & \multicolumn{3}{c|}{\textbf{EmoDB (de)}}                                                                     & \multicolumn{3}{c|}{\textbf{CaFE (fr)}}                                                                      & \multicolumn{3}{c|}{\textbf{EMOVO (it)}}                                                                     & \multicolumn{3}{c}{\textbf{MESD (es)}}                                                                       \\ \cline{6-17} 
\textit{\textbf{Idx}} & \textit{\textbf{HR}}$^{\mathrm{a}}$ & \textit{\textbf{TR}}$^{\mathrm{b}}$ & \textit{\textbf{EX}}$^{\mathrm{c}}$ & \textit{\textbf{BO}}$^{\mathrm{d}}$ & \textit{\textbf{UA(\%)$\uparrow$}} & \textit{\textbf{WA(\%)$\uparrow$}} & \textit{\textbf{F1(\%)$\uparrow$}} & \textit{\textbf{UA(\%)$\uparrow$}} & \textit{\textbf{WA(\%)$\uparrow$}} & \textit{\textbf{F1(\%)$\uparrow$}} & \textit{\textbf{UA(\%)$\uparrow$}} & \textit{\textbf{WA(\%)$\uparrow$}} & \textit{\textbf{F1(\%)$\uparrow$}} & \textit{\textbf{UA(\%)$\uparrow$}} & \textit{\textbf{WA(\%)$\uparrow$}} & \textit{\textbf{F1(\%)$\uparrow$}} \\ \hline
1                     &                      &                      &                      &                      & 84.97                              & 83.79                              & 84.04                              & 67.33                              & 66.72                              & 66.02                              & 49.90                              & 49.90                              & 44.22                              & 64.75                              & 65.18                              & 64.33                              \\
2                     & $\checkmark$         &                      &                      & $\checkmark$         & 83.49                              & 83.26                              & 83.02                              & 67.26                              & 67.30                              & 66.87                              & 52.28                              & 52.28                              & 48.31                              & 62.31                              & 62.95                              & 62.03                              \\
3                     & $\checkmark$         & $\checkmark$         &                      & $\checkmark$         & 79.53                              & 79.50                              & 78.90                              & 63.76                              & 63.37                              & 62.82                              & \textbf{57.54}                     & \textbf{57.54}                     & \textbf{53.68}                     & 58.83                              & 58.85                              & 58.40                              \\
4                     & $\checkmark$         & $\checkmark$         & $\checkmark$         &                      & 72.61                              & 71.56                              & 71.30                              & 52.58                              & 53.30                              & 51.73                              & 53.17                              & 53.17                              & 50.03                              & 55.57                              & 56.15                              & 55.35                              \\
5                     & $\checkmark$         & $\checkmark$         & $\checkmark$         & $\checkmark$         & \textbf{85.94}                     & \textbf{85.17}                     & \textbf{85.27}                     & \textbf{68.32}                     & \textbf{68.29}                     & \textbf{67.74}                     & 55.56                              & 55.56                              & 51.04                              & \textbf{64.87}                     & \textbf{65.21}                     & \textbf{64.65}                     \\ \hline
\multicolumn{17}{l}{$^{\mathrm{a}}$\textbf{HR}: use external dataset of high recourse, $^{\mathrm{b}}$\textbf{TR}: translation, $^{\mathrm{c}}$\textbf{EX}: expressivity of translation, $^{\mathrm{d}}$\textbf{BO}: proposed bootstrapping data selection process}
\end{tabular}
\vspace{-12.5pt}
\label{tab:ablation}
\end{center}
\end{table*}
}
{\begin{table*}[]
\setlength\tabcolsep{5.25pt}
\renewcommand{\arraystretch}{1.15}
\caption{Experiment to develop guidelines for selecting source dataset}
\vspace{-10pt}
\begin{center}
\begin{tabular}{p{50pt}|ccc|ccc|ccc|ccc}
\hline
\multicolumn{1}{c|}{\textbf{}}          & \multicolumn{3}{c|}{\textbf{EmoDB (de)}}                                                                     & \multicolumn{3}{c|}{\textbf{CaFE (fr)}}                                                                      & \multicolumn{3}{c|}{\textbf{EMOVO (it)}}                                                                     & \multicolumn{3}{c}{\textbf{MESD (es)}}                                                                       \\ \cline{2-13} 
\multicolumn{1}{c|}{\textit{\textbf{}}} & \textit{\textbf{UA(\%)$\uparrow$}} & \textit{\textbf{WA(\%)$\uparrow$}} & \textit{\textbf{F1(\%)$\uparrow$}} & \textit{\textbf{UA(\%)$\uparrow$}} & \textit{\textbf{WA(\%)$\uparrow$}} & \textit{\textbf{F1(\%)$\uparrow$}} & \textit{\textbf{UA(\%)$\uparrow$}} & \textit{\textbf{WA(\%)$\uparrow$}} & \textit{\textbf{F1(\%)$\uparrow$}} & \textit{\textbf{UA(\%)$\uparrow$}} & \textit{\textbf{WA(\%)$\uparrow$}} & \textit{\textbf{F1(\%)$\uparrow$}} \\ \hline
baseline                                & 84.97                              & 83.79                              & 84.04                              & 67.33                              & 66.72                              & 66.02                              & 49.90                              & 49.90                              & 44.22                              & 64.75                              & 65.18                              & 64.33                              \\
w/o soft label$^{\mathrm{a}}$                          & 85.07                              & 84.44                              & 84.55                              & 67.79                              & 68.00                              & 66.78                              & 54.56                              & 54.56                              & 50.14                              & 62.66                              & 63.02                              & 62.37                              \\
BIIC-Podcast                            & 81.95                              & 81.83                              & 81.72                              & 64.68                              & 64.18                              & 62.80                              & \textbf{55.66}                     & \textbf{55.66}                     & \textbf{51.99}                     & 62.31                              & 62.64                              & 62.17                              \\
default                                 & \textbf{85.94}                     & \textbf{85.17}                     & \textbf{85.27}                     & \textbf{68.32}                     & \textbf{68.29}                     & \textbf{67.74}                     & 55.56                              & 55.56                              & 51.04                              & \textbf{64.87}                     & \textbf{65.21}                     & \textbf{64.65}                     \\ \hline
\multicolumn{13}{l}{$^{\mathrm{a}}$ Using \eqref{eq:1} as selecting criterion}
\end{tabular}
\vspace{-20pt}
\label{tab:other}
\end{center}
\end{table*}
}

\vspace{-2pt}
\subsection{Ablation study}
In this section, we conduct an ablation study using emotion2vec as the upstream model and our default selection criterion \eqref{eq:2} to validate each component of our method.
We break down our approach into four key components: training on external data from a high-resource language, translation, translation with expressivity, and bootstrapping data selection process.
As shown in Table~\ref{tab:ablation}, all components are necessary for achieving optimal and consistent performance.
Specifically, comparing row $4$ and $5$, we observe that bootstrapping data selection plays a crucial role in achieving significant performance improvement across all languages.
We also find that bootstrapping consistently improves performance across all ablation settings.
Therefore, results with bootstrapping are presented in row $2$ and $3$ of the ablation table.

By comparing row $3$ and $5$, it is clear that our method achieves much better overall performance, indicating that the expressivity of the synthesized training data is important.
This finding is unsurprising, as most emotional information is conveyed through paralinguistic details in speech.
The significant difference between these settings aligns with the strong performance of the expressivity encoder and supports our hypothesis regarding the effectiveness of our method.
The $D_{syn}$ in row $3$ was generated by SeamlessM4T-Large v2, the non-expressive version of Seamless Expressive.
Without the expressivity encoder, it fails to synthesize data that benefits emotion recognition.

Interestingly, when using the source dataset without translation (row $2$), the audio retains its paralinguistic information.
Based on this and the results in row $2$ and $3$, we conclude that expressivity may be more important than training on the target language itself.
However, both of these settings are still inferior to the baseline, and only when the external data is both expressive and matched to the target language is a notable performance boost observed.

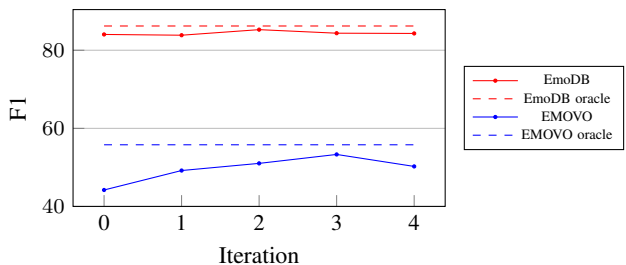
\begin{figure}
\centering
\begin{tikzpicture}
\begin{axis}[  
    xtick pos=lower,
    xlabel=Iteration,
    ylabel=F1,
    ylabel near ticks,
    xlabel near ticks,
    yticklabel style = {font=\small},
    xticklabel style = {font=\small},
    ytick pos=left, symbolic x coords={0, 1, 2, 3, 4}, enlargelimits=0.1, xtick=data, grid=major,
    xmajorgrids=false, width=0.36\textwidth,
    height=4.2cm,
    legend style={at={(1.05,0.5)},anchor=west, nodes={scale=0.6, transform shape}},
]
\addplot[mark=*, mark options={scale=0.3}, color=red] coordinates {(0, 84.04266667) (1, 83.847) (2, 85.27066667) (3, 84.363) (4, 84.29966667)};
\addplot[dashed, mark=none, color=red] coordinates {(0, 86.208) (1, 86.208) (2, 86.208) (3, 86.208) (4, 86.208)};
\addplot[mark=*, mark options={scale=0.3}, color=blue] coordinates {(0, 44.22333333) (1, 49.20633333) (2, 51.037) (3, 53.32333333) (4, 50.25633333)};
\addplot[dashed, mark=none, color=blue] coordinates {(0, 55.797) (1, 55.797) (2, 55.797) (3, 55.797) (4, 55.797)};
\legend{EmoDB, EmoDB oracle, EMOVO, EMOVO oracle}
\end{axis}
\end{tikzpicture}
\vspace{-10pt}
\caption{Effect of different numbers of iterations for bootstrapping on EmoDB and EMOVO. The dashed line is the oracle result where we select the best step in each fold and seed.}
\label{fig:steps}
\vspace{-15pt}
\end{figure}

\subsection{Selecting proper source dataset}
We conducted two additional experiments about the source dataset $D_{src}$ to evaluate the impact of using different source datasets and to develop guidelines for selecting $D_{src}$.
The first experiment investigates the effectiveness of our method when the assumption of access to soft labels for the source data does not hold—in other words, when only one-hot label is available for the source dataset.
In this case, our default selection criterion \eqref{eq:2} is no longer applicable, so we use \eqref{eq:1} as the selection criterion.

As shown in Table~\ref{tab:other}, we still observe performance gains in $3$ out of $4$ languages; however, the improvements are less substantial compared to using our default selection criterion.
The advantage of soft labels likely stems from the complex, multidimensional nature of human emotion.
When using criterion \eqref{eq:1}, false positive data (where the prediction matches its highest-voted class but actually comes from a significantly different distribution) may be selected, leading to diminished performance.
In contrast, selecting based on KL divergence, as in \eqref{eq:2}, allows us to capture more nuanced differences, such as distinguishing between sadness with happiness versus sadness with anger.
Consequently, we are more likely to select data that aligns better with the distribution of the target dataset $D_{tgt}$.

In the second experiment, we switch the source dataset from MSP-Podcast to BIIC-Podcast to assess the effect of changing the source language $L_s$.
As shown in Table~\ref{tab:other}, using BIIC-Podcast as the source dataset generally leads to performance degradation.
We suspect this may be related to how emotions are expressed and the level of expressivity in different languages.
Previous studies have found that Chinese, for example, tends to be less expressive than English, or that emotions are conveyed through more subtle paralinguistic cues, which are harder to capture \cite{app8122629}. 
Subtle paralinguistic cues may be difficult for the S2ST to detect and capture.
As we observed in Tab~\ref{tab:ablation}, without expressivity, even with bootstrapping data selection, $D_{syn}$ may still lead to performance degradation.
This could negatively affect the quality of synthesized data.

Based on these results, we derive two guidelines for selecting an appropriate source dataset $D_{src}$:
\begin{itemize}
    \item Although $D_{src}$ with simple one-hot annotations can provide some improvement, it is preferable to select a dataset with soft labels, as these are beneficial in the bootstrapping data selection process.
    \item Choosing datasets in more expressive languages may help the data synthesis process preserve more emotional information, thus improving the quality of $D_{syn}$ and the performance.
\end{itemize}

\subsection{Number of iterations for bootstrapping}
In all experiments, we fix the number of iterations to $2$, based on empirical findings that this configuration achieves a good balance between performance and training time.
However, the optimal number of iterations may vary depending on the upstream models and target datasets.
Fig.~\ref{fig:steps} shows the performance for different numbers of iterations alongside the oracle result (where the best number of iterations is selected for each fold and seed) for emotion2vec on the EmoDB and EMOVO datasets.

As shown in Fig.~\ref{fig:steps}, the optimal number of iterations is $2$ for EmoDB and $3$ for EMOVO. 
While $2$ iterations already provide a substantial performance boost, we observe an additional gain of $2.3$ F1 points with one more iteration.
Furthermore, a significant gap remains between the fixed number of iterations and the oracle result, indicating room for improvement.
These findings suggest that an adaptive method for dynamically determining whether to continue or stop the bootstrapping process is needed.
We leave this exploration for future work.

\section{Conclusion}
In this paper, we demonstrate that multilingual SER can be significantly improved by leveraging large-scale datasets in the target language.
We employ an expressive S2ST model and propose a simple yet effective data selection process to construct a synthesized dataset, leading to substantial performance gains.
Our experiments show the generalizability of the proposed method across various upstream models and datasets in different languages.
Additionally, we provide guidelines for selecting an appropriate source dataset to further enhance performance.
We hope that our work paves the way for more robust and scalable approaches in multilingual SER.

\section{Limitation and Future work}
Although our proposed method is effective and generalizable across different upstream models and languages, there are still limitations that can be addressed. Currently, we treat the number of iterations for bootstrapping as a fixed hyperparameter, and an adaptive strategy for determining the optimal number of iterations is needed for further improvement.
Additionally, achieving the highest performance gains requires access to soft labels from the source SER dataset. 
These factors suggest that there is still room to develop a pipeline that can be applied across all scenarios.

\section{Acknowledgement}
We thank to National Center for High-performance Computing (NCHC) of National Applied Research Laboratories (NARLabs) in Taiwan for providing computational and storage resources.

\bibliographystyle{IEEEtranS}
\bibliography{IEEEabrv, ref}

\end{document}